\documentclass{ws-procs961x669}            
\begin{document}
\title{On explaining prompt emission from GRB central engines with photospheric emission model}

\author{M. Bhattacharya$^{1*}$ and P. Kumar$^{2}$}


\address{${}^1$Department of Physics, Pennsylvania State University, University Park, PA 16802, USA\\
$^*$E-mail: mmb5946@psu.edu\\
${}^2$Department of Astronomy, University of Texas, Austin, TX 78712, USA}

\begin{abstract}
Although the observed spectra for gamma-ray burst (GRB) prompt emission is well constrained, the underlying radiation mechanism is still not very well understood. We explore photospheric emission in GRB jets by modelling the Comptonization of fast cooled synchrotron photons whilst the electrons and protons are accelerated to highly relativistic energies by repeated energy dissipation events as well as Coulomb collisions. In contrast to the previous simulations, we implement realistic photon-to-particle number ratios of $N_{\gamma}/N_e \sim 10^{5}$ or higher, that are consistent with the observed radiation efficiency of relativistic jets. Using our Monte Carlo radiation transfer (MCRaT) code, we can successfully model the prompt emission spectra when the electrons are momentarily accelerated to highly relativistic energies (Lorentz factor $\sim 50-100$) after getting powered by $\sim30-50$ episodic dissipation events in addition to their Coulomb coupling with the jet protons, and for baryonic outflows that originate from moderate optical depths $\sim20-30$. We also show that the resultant shape of the photon spectrum is practically independent of the initial photon energy distribution and the jet baryonic energy content, and hence independent of the emission mechanism.
\end{abstract}

\keywords{Gamma-ray burst: general; methods: numerical; radiation mechanisms: thermal; radiative transfer; scattering.}

\bodymatter

\section{Introduction}
\label{sec1}
The radiation mechanism responsible for the prompt emission of long-duration GRBs has remained elusive ever since their discovery. The observed photon spectrum has a distinctly non-thermal shape and is generally modelled using the Band function \cite{Band93}, which is a smoothly connected broken power-law with peak energy $E_{\rm peak} \sim 300\, {\rm keV}$ and the low/high energy dependence \cite{Preece00,Kaneko06} given by $f_{\nu} \propto \nu^0/f_{\nu} \propto \nu^{-1.2}$. A robust radiation mechanism should explain the non-thermal features observed in the prompt emission spectrum in a self-consistent manner. The two most widely explored models to this end are the synchrotron and photospheric models \cite{RM94,Piran04,KZ15}. 

In the synchrotron model, the electrons accelerated by either internal shocks \cite{RM94} or magnetic reconnection \cite{Gia06} produce the prompt radiation via synchrotron emission process \cite{Piran99}. Although this model accounts for the non-thermal nature of the photon spectrum, it cannot explain the high observed radiation efficiencies up to few tens of percent \cite{Zhang07}. Moreover, the observed hard GRB spectra at low energies cannot be explained with synchrotron emission process \cite{Preece98,Ghirlanda03}. Owing to these shortcomings of the synchrotron model, many researchers have considered photospheric emission model in more detail \cite{RM05,MBh20}. Unlike the synchrotron model, photospheric model naturally explains the observed high radiation efficiencies and spectral shape is completely determined by photon-matter interactions irrespective of the dissipation mechanism involved. Although there have been many successful attempts to explain the high-energy non-thermal behaviour using sub-photospheric dissipation \cite{Gia06,LB10,MBh18}, explaining the low-energy non-thermal tails has turned out to be really challenging \cite{CL15}. 

Here we study the sub-photospheric Comptonisation of synchrotron seed photons \cite{Granot00,Piran04} to find the plausible conditions under which both low/high-energy non-thermal behaviour and observed peak energy can be explained self-consistently. The particles in the jet are accelerated to relativistic energies by repeated dissipation events such as internal shocks \cite{RM94,Toma11} or magnetic reconnection \cite{Thompson94,Gia06}. The photons continue to scatter electrons while the outflow is optically thick and gain energy until either the average photon energy matches that of the electrons or the outflow becomes optically thin so that the photons escape the photosphere. We determine a correlation between the number of reheating events and initial optical depth, and perform an exhaustive parametric space search to obtain a Band-like photon spectrum. We also perform analytical calculations to examine the evolution of photon energy spectrum with multiple scatterings and validate the numerical results.

\section{Energy requirement for photons and particles}
We first estimate the average energy $E_{\gamma,\rm avg}$ that the jet photons need to have to generate a Band-like output spectrum. As most of this energy is transferred by hot electrons through Comptonisation, this also places a threshold energy requirement $\gamma_{e,\rm crit}$ on the electrons. As the electrons are maintained at $\gamma_{e} \sim \gamma_{e,crit}$ by energy gained from either Coulomb collisions or repeated dissipation events and subsequent cooling due to IC, we further constrain the injected energy and the number of episodic dissipation events required.

\subsection{Analytical estimate for $E_{\gamma,\rm avg}$ and $\gamma_{e,\rm crit}$}
The observed photon spectrum has a Band-like shape with a low/high-energy dependence, $f_{\nu} \propto \nu^0/\nu^{-1.2}$ in the energy range $\sim 10\ {\rm keV} - 300\ {\rm keV}$/$\sim 300\ {\rm keV} - 10\ {\rm MeV}$, where $f_{\nu}$ denotes the photon flux per unit frequency. The average observed energy of each photon in the lab frame is then $E_{\gamma,\rm avg}^{\prime} \sim 300\, {\rm eV}$, assuming a jet bulk Lorentz factor $\Gamma = 300$. Equating the total energy content of the photons with the maximum energy that the electrons can deposit via Comptonisation gives 
\begin{equation}
N_{\gamma}E_{\gamma,{\rm avg}}^{\prime} = N_{e}(\gamma_{e,{\rm crit}} - 1)m_{e}c^{2} \left(\tau_{{\rm in}}t_{{\rm dyn}}^{\prime}/t_{{\rm IC}}^{\prime}\right), \\
\label{ge_no_shock_with_supCoul}
\end{equation}
where $\tau_{in}$ is the initial optical depth, $t_{{\rm dyn/IC}}^{\prime}$ is the dynamical/inverse-Compton (IC) timescale and $\tau_{{\rm in}}t_{{\rm dyn}}^{\prime}/t_{{\rm IC}}^{\prime} \sim$ number of times the electrons interact with photons during jet expansion. The characteristic dynamical timescale is $t_{\rm dyn}^{\prime} = R_{\rm in}/\Gamma c$ while the IC timescale is $t_{\rm IC}^{\prime} = 3(\gamma_e - 1)m_e c/4U_{\gamma}^{\prime}\sigma_T \gamma_e^2 \beta_e^2$. Here, $R_{\rm in}$ is the photon injection radius, $\sigma_T$ is the Thomson cross section, $\gamma_e$ is the electron energy and $U_{\gamma}^{\prime}$ is the radiation energy density. Substituting typical GRB parameters gives $\gamma_{e,crit} = 1.352$. 

We now estimate the electron super-Coulomb efficiency parameter $\eta$ which leads to $\gamma_{e} = 1.352$. For electrons to remain in equilibrium as a result of Coulomb heating and IC cooling over the jet expansion timescale, 
\begin{equation}
\frac{(\gamma_{e} - 1)m_{e}c^{2}}{5\times10^{-19}n_{e}^{\prime}}.\ \frac{(8.3\times10^{-15}T_{e}^{\prime 3/2} + \beta_{p}^{3})}{\beta_{p}^{2}}.\ \frac{1}{\eta} = \frac{3}{4}\frac{(\gamma_{e} - 1)m_{e}c}{U_{\gamma}^{\prime}\sigma_{T}\gamma_{e}^{2}\beta_{e}^{2}},
\label{eqn_super_Coul}
\end{equation}
where electron density $n_{e}^{\prime} = L/(4\pi R^2 m_{p}c^{3} \Gamma^{2}) = 4.17\times10^{15}\ {\rm cm^{-3}}$, radiation energy density $U_{\gamma}^{\prime} = L_{\gamma}/(4\pi R^{2}\Gamma^{2}c) = 2\times10^{11}\ {\rm erg/cm^{3}}$ and $T_{e}^{\prime} = \frac{1}{k_{B}}(\gamma_{e,ad} - 1)(\gamma_{e} - 1)m_{e}c^{2} = 1.98\times10^{9}\left(\gamma_{e} - 1/\gamma_{e}\right)$ is the electron temperature for a Maxwellian distribution. Here, $\gamma_{e,ad} = (4\gamma_e +1)/(3\gamma_e)$ is the adiabatic index of the electrons and $\beta_p$ is the speed of protons divided by the speed of light. Substituting $\gamma_{p} \sim 1.123$ for $\tau_{in}=8$ and $\gamma_{e} = \gamma_{e,crit}= 1.352$, we obtain $\eta = 4.5$ \cite{MBh18}. 

The electrons cannot be continuously heated by Coulomb collisions if the protons cool down to energies comparable to that of electrons within the dynamical timescale. While the protons cool down due to Coulomb collisions and adiabatic expansion, the electrons gain energy through Coulomb and get cooled due to adiabatic cooling and IC. The electrons cannot be heated any further when
\begin{eqnarray}
N_{p}(\gamma_{p}-1)m_{p}c^{2} - N_{p}\int_{0}^{t^{\prime}}\frac{5\times10^{-19}n_{e}^{\prime}\beta_{p}^{2}}{[0.73(\gamma_{e} - 1/\gamma_{e})^{3/2} + \beta_{p}^{3}]}\eta dt^{\prime} 
- N_{p}\int_{0}^{t^{\prime}}\frac{(\gamma_{p}-1)m_{p}c^{2}}{R/\Gamma c}dt^{\prime} \nonumber \\
= N_{e}(\gamma_{e}-1)m_{e}c^{2} + N_{e}\int_{0}^{t^{\prime}}\frac{5\times10^{-19}n_{e}^{\prime}\beta_{p}^{2}}{[0.73(\gamma_{e} - 1/\gamma_{e})^{3/2} + \beta_{p}^{3}]}\eta dt^{\prime} \nonumber \\
- N_{e}\int_{0}^{t^{\prime}}\frac{(\gamma_{e}-1)m_{e}c^{2}}{R/\Gamma c}dt^{\prime} - N_{e}\int_{0}^{t^{\prime}}(4/3)U_{\gamma}^{\prime}\sigma_{T}(\gamma_{e}^{2} - 1)cdt^{\prime}.\ \ 
\end{eqnarray}
which simplifies for a neutral jet to, $1.5\times10^{-3}(1 - {\rm ln}\lambda) + 3.21\times10^{-5}(1 - 1/\lambda) = 1.15\times10^{-4}\eta(1 - 1/\lambda)\tau_{in}$. For $\lambda \sim 2$ i.e. for protons to cool down to electron energies in $t^{\prime} = 2 t_{dyn}^{\prime}$, $\eta \tau_{in} \sim 8.28$. This means that for $\tau_{in} \gtrsim 10$, the protons cool down too fast and super-Coulomb interaction cannot keep the electrons hot beyond $t^{\prime} = 2 t_{dyn}^{\prime}$. This necessitates the heating of electrons by some alternate sub-photospheric dissipation mechanism. Even though the electrons tend to cool down rapidly due to Comptonization, their energy can still be maintained at $\gamma_e \gtrsim \gamma_{e,crit}$ provided the episodic heating events are frequent enough.

\subsection{Electron heating by repeated dissipation events}
As the total energy gained by the photons is the total energy that is transferred by the electrons through Comptonisation
\begin{eqnarray}
N_{\gamma}\tau_{in}^{2/3}\left(E^{\prime}_{\gamma,avg,obs} - E^{\prime}_{\gamma,avg,i}\right) = N_{e}\int_{0}^{\tau_{in}t_{dyn}^{\prime}}(4/3)U_{\gamma}^{\prime}\tau_{in}^{-2/3}\sigma_{T}\gamma_{e}^{2}\beta_{e}^{2}c\ dt^{\prime}.\nonumber
\end{eqnarray}
which simplifies to 
\begin{eqnarray}
10^{5}\tau_{in}^{2/3}E^{\prime}_{\gamma,avg,obs} = 8.72\times10^{6}L_{\gamma,50}\tau_{in}^{-2/3}\int_{R_{in}}^{\tau_{in}R_{in}}(\gamma_{e}^{2} - 1)\frac{dR}{R^2} \nonumber \\
\sim \frac{8.72\times10^{6}L_{\gamma,50}\tau_{in}^{-2/3}(\gamma_{e}^{2} - 1)}{R_{in}}.\nonumber
\end{eqnarray}
Substituting $R_{in} \sim 2.17\times10^{11}\ \tau_{in}^{-1}$ cm, yields $\gamma_{e,crit} \sim \sqrt{1 + 1.2\ L_{\gamma,50}^{-1}\ \tau_{in}^{1/3}}$. The critical electron energy obtained here is similar to $\gamma_{e,crit} \sim 1.352$ obtained earlier, especially for small initial optical depths $\tau_{in} \sim 1$. 

We consider repeated sub-photospheric dissipation events that can re-accelerate the electrons as well as protons to their initial energies. 

Using the fact that the electrons remain in equilibrium with energy $\gamma_{e,\rm crit} \sim \sqrt{1 + 1.2\ L_{\gamma,50}^{-1}\ \tau_{\rm n}^{1/3}}$, we constrain the energy injected per electron $E_{inj} = N_{rh}(\gamma_{e,in} - 1)m_e c^2$. In particular,
\begin{eqnarray}
N_{e}\tau_{in}^{4/3}(\gamma_{e} - 1)m_{e}c^{2} = N_{e}\int_{0}^{\tau_{in}t_{dyn}^{\prime}}\frac{5\times10^{-19}n_{e}^{\prime}\beta_{p}^{2}\tau_{in}^{-4/3}}{[0.73(\gamma_{e} - 1/\gamma_{e})^{3/2} + \beta_{p}^{3}]} dt^{\prime}\nonumber \\
+ N_{e}N_{rh}(\gamma_{e,in} - 1)m_{e}c^{2} - N_{e}\int_{0}^{\tau_{in}t_{dyn}^{\prime}}(4/3)U_{\gamma}^{\prime}\tau_{in}^{-2/3}\sigma_{T}\gamma_{e}^{2}\beta_{e}^{2}c\ dt^{\prime},
\label{tau_Nrh1}
\end{eqnarray}
where $\gamma_e \approx \gamma_{e,crit}$ and $\tau_{in}^{-4/3}$/$\tau_{in}^{-2/3}$ is the adiabatic cooling factor for relativistic electrons/photons. We consider the episodic dissipation events to be equally spaced over the jet expansion timescale $\tau_{in}t_{dyn}^{\prime}$ and to supply fixed energy (equal to initial energy, $\gamma_{e,in}$/$\gamma_{p,in}$) to the electrons/protons at each instance. Substituting $R_{\rm in} = 2.17\times10^{11}\ \tau_{\rm in}^{-1}\ {\rm cm}$ and simplifying yields
\begin{eqnarray}
\tau_{in}^{4/3}(\gamma_e - 1)m_e c^2 = \frac{10^{-4}\tau_{in}^{-4/3}}{[0.73(\gamma_{e} - 1/\gamma_{e})^{3/2} + (2/\tau_{in})^{3/2}]} 
+ E_{\rm inj,cr} - 3.21\times 10^{-5}\tau_{in}^{-2/3},
\label{Nrh_ad}
\end{eqnarray}
which constrains the critical injected energy $E_{inj,cr}(\tau_{in}) = N_{rh}(\gamma_{e,in} - 1)m_{e}c^{2}$ per electron in terms of $\tau_{in}$. Equation (\ref{Nrh_ad}) is only a necessary and not sufficient condition to obtain Band-like photon spectrum as it determines the average photon energy but does not impose any constraints on the general shape of the photon spectrum. For large $E_{inj}$, photon peak energy $E_{\gamma,peak} \gg E_{\gamma,obs} \sim 300\ {\rm keV}$ while $E_{\gamma,peak} \ll E_{\gamma,obs}$ for large $\tau_{in}$, due to significant energy loss from adiabatic cooling.

\section{MCRaT code description}
We discuss here the implementation of our MCRaT code and give an overview of the basic physics included. We first list the jet parameters and describe the initial distributions of the particles and the photons. We then discuss how the jet particles and photons are affected by the physical processes. Lastly, we describe the algorithm of our photospheric MCRaT code.

\subsection{Relativistic jet parameters}
The jet parameters provided as input for the code are:  
\begin{itemlist}
\item{{\it Isotropic equivalent luminosity,} $L$}: The bulk of the jet luminosity is contributed by the protons. We consider $L = 10^{51}, 10^{52}\ {\rm erg/s}$ \cite{Liang07}. 

\item{{\it Jet bulk Lorentz factor,} $\Gamma$}: The bulk Lorentz factor is related to $L$ and $E_{\gamma,peak}$. We consider $\Gamma = 30, 100, 300$ in this work \cite{Xue09}.

\item{{\it Initial optical depth,} $\tau_{\rm in}$}: The optical depth is measured relative to $R = L\sigma_T/(8\pi m_p c^3 \beta \Gamma^3 \tau)$ which is the photon radial distance from central engine in the observer frame. $\tau_{\rm in}$ corresponds to the radial distance from central engine where all particles and photons are injected. We consider $\tau_{\rm in} = 10, 20, 40$. 
\end{itemlist}

\subsection{Initial distributions}
We describe here the initial energy and velocity distributions of the jet electrons, protons and photons. 

\begin{itemlist}
\item{{\it Electrons and protons:}} We consider a charge-neutral jet with particle number $N_{e} = N_{p} = 2\times 10^{2}$. The initial particle velocities are distributed randomly in the jet-comoving frame. All particles are uniformly distributed in the jet-comoving frame at initial time. The initial energy of the electrons are determined from the Maxwellian distribution with temperature $k_{B}T_{e,in}^{\prime} = (\gamma_{e,ad,in}-1)(\gamma_{e,in} - 1)m_e c^2$ while the protons are mono-energetic with $\gamma_{p} = \gamma_{p,in}$. We consider $\gamma_{e,in} = 2, 10, 30, 100$ and $\gamma_{p,in} = 1.01,1.1$ for our simulations. 

\item{{\it Photons:}} In order to maintain $N_{\gamma}/N_{e} = 10^{5}$, we consider $N_{\gamma} = 2\times10^{7}$ for our simulations \cite{MBh18}. The initial photon velocities are randomly distributed in the jet-comoving frame and photon positions are uniformly distributed within a cone of solid angle $1/\Gamma$ pointing towards the observer. The initial photon energy distribution is given by the synchrotron distribution for fast cooling electrons \cite{Piran04}
\begin{equation}
\label{ph_seed}
f_{\nu} = \left\{
\begin{array}{ll}
\left(\frac{\nu_{ac}}{\nu_{sa}}\right)^{11/8}\:\left(\frac{\nu}{\nu_{ac}}\right)^{2}, & \nu_{min} < \nu < \nu_{ac}\\
\left(\frac{\nu}{\nu_{sa}}\right)^{11/8}, & \nu_{ac} < \nu < \nu_{sa} \\
\left(\frac{\nu}{\nu_{sa}}\right)^{-1/2}, & \nu_{sa} < \nu < \nu_{m} \\
\left(\frac{\nu_{m}}{\nu_{sa}}\right)^{-1/2}\:\left(\frac{\nu}{\nu_{m}}\right)^{-p/2}, & \nu_{m} < \nu < \nu_{max}\\
\end{array}
\right. 
\end{equation}
where $f_{\nu}$ is the peak normalised photon flux per unit frequency and $p = 2.5$ is the spectral index at high energies \cite{KZ15}. 
\end{itemlist}

\subsection{Physical processes in the jet}
Here we discuss the physical interactions between the electrons, protons and photons that can further affect the output photon spectrum. 

\begin{itemlist}
\item{{\it Adiabatic cooling:}} As the relativistic jet expands outward, the photon and particle energies drop considerably as  
\begin{eqnarray}
(\gamma_{e,f} - 1)/(\gamma_{e,i} - 1) = \left(R_{\rm in} + \beta c \Gamma t^{\prime}_{f}/R_{\rm in} + \beta c \Gamma t^{\prime}_{i}\right)^{-2(\gamma_{\rm ad,e} - 1)},\nonumber \\
(\gamma_{p,f} - 1)/(\gamma_{p,i} - 1) = \left(R_{\rm in} + \beta c \Gamma t^{\prime}_{f}/R_{\rm in} + \beta c \Gamma t^{\prime}_{i}\right)^{-2(\gamma_{\rm ad,p} - 1)},\nonumber \\
E_{\gamma,f}/E_{\gamma,i} = \left(R_{\rm in} + \beta c \Gamma t^{\prime}_{f}/R_{\rm in} + \beta c \Gamma t^{\prime}_{i}\right)^{-2/3},
\end{eqnarray}
where the subscript $i/f$ denotes the initial/final value of the physical quantity and $\gamma_{ad,e/p} = (4\gamma_{e/p} +1)/(3\gamma_{e/p})$ is the electron/proton adiabatic index. 

\item{{\it Coulomb collisions:}} While the electrons are continuously heated by the protons in the jet, they also interact with each other to quickly attain thermal equilibrium. The corresponding timescales are given by \cite{Schlickheiser02}
\begin{eqnarray}
\dot{E}_{e-p} = \frac{5\times10^{-19}n_{e}^{\prime}\beta_{\rm p,avg}^{2}}{8.3\times10^{-15}T_{\rm e,avg}^{\prime 3/2} + \beta_{\rm p,avg}^{3}},\nonumber \\
\dot{E}_{e-e} = \frac{5\times10^{-19}n_{e}^{\prime}\beta_{\rm e,avg}^{2}}{8.3\times10^{-15}T_{\rm e,avg}^{\prime 3/2} + \beta_{\rm e,avg}^{3}},
\label{Coul}
\end{eqnarray}
where $\beta_{p,avg}$, $\beta_{e,avg}$ and $T_{e,avg}^{\prime}$ are number-averaged quantities. The electron distribution is re-initialized to Maxwellian distribution on a timescale $t_{e-e}^{\prime} = (\gamma_{e,avg}-1)m_e c^2/\dot{E}_{e-e} \ll t_{e-p}^{\prime}$. 

\item{{\it IC scattering:}} The distance $s^{\prime}$ that a photon travels before scattering an electron is given by probability density $p(s^{\prime}) \propto {\rm exp}(-s^{\prime}/l_{\rm mfp}^{\prime})$, where $l_{\rm mfp}^{\prime} = 1/(n_{e}^{\prime}\sigma_T)$ is the photon mean free path. The scattering probability of a particular electron with a photon is \cite{MBh18}
\begin{equation}
P_{sc}(\beta_{e},\theta_{e}^{\prime}) = \frac{1}{4\pi \beta_e^{2}}(1 - \beta_{e} {\rm cos}\ \theta_{e}^{\prime}),
\label{Psc}
\end{equation} 
where $\beta_{e}$ is the electron speed divided by speed of light and $\theta_{e}^{\prime}$ is the angle between electron and photon velocities before scattering. Average number of scatterings that a photon experiences before escaping is $\sim 2\tau_{in}$ \cite{Begue13}. 

\item{{\it Pair production/annihilation:}} Due to the episodic jet dissipation events, the electrons are often accelerated to highly relativistic energies $\gamma_{e} = \gamma_{e,in} \sim 100$ and can scatter energetic photons with $E_{\gamma}^{\prime} \gtrsim 10\ E_{\gamma,peak}^{\prime}$ to energies $\gtrsim 4 E_{\gamma,avg}^{\prime}\gamma_{e}^{2} \sim m_e c^2 \sim 5\times10^{5}\ {\rm eV}$, before cooling down rapidly to non-relativistic energies. For synchrotron photons that we consider (see equation \ref{ph_seed}), a considerable fraction $\sim 30\%$ have sufficient energy to generate electron-positron pairs which can affect the shape of the output photon spectrum, especially for large $\tau_{in} \gtrsim 10$. 
\end{itemlist}

\subsection{Code implementation}
The travel distances are first drawn for all photons depending on their mean free path and the photons are propagated. The new photon positions are evaluated to check if any photon escapes the photosphere, in which case the energy is calculated and stored. Other photons are stored in a priority queue where they are ordered based on travel distances. Next, the photon at the top of the queue is propagated, a proton is randomly selected and an electron is selected using the scattering probability, $P_{sc}$. The particle and photon energies are then updated due to adiabatic cooling and Coulomb collisions. The outgoing velocities and energies of the photon and electron are calculated if IC scattering occurs. Subsequently, the next photon in the queue is drawn and electron-positron pair production cross section is evaluated. If the cross section is large, a new electron and positron are generated and photons are not placed back in the queue. If positron number is non-zero, a positron is drawn randomly and the pair annihilation cross section with the electron is calculated. Two new photons are created and added to the queue, if the cross section is significant. The method described above is repeated until a third of the total photons in the jet escape and a time-averaged output photon spectrum is obtained.

\section{Photospheric simulation results}
\label{sim_results}
Here we present the results of our photospheric MCRaT simulations. The photon energy spectrum and the electron kinetic energy spectrum are shown in the lab frame at the end of each simulation in all the figures. The photon and electron energy spectrum are Doppler boosted from the jet-comoving frame to the lab frame in all the figures. 
The electron kinetic energy spectra are peaked at significantly larger energies compared to the photon spectra for all the simulations as shown in the figures.
In the rest of this paper, we denote the low/high energy photon spectral index by $\alpha$/$\beta$ and the observed photon peak energy by $E_{\gamma,obs}$.

\begin{figure}[h]%
\begin{center}
  \parbox{2.4in}{\includegraphics[width=2.4in]{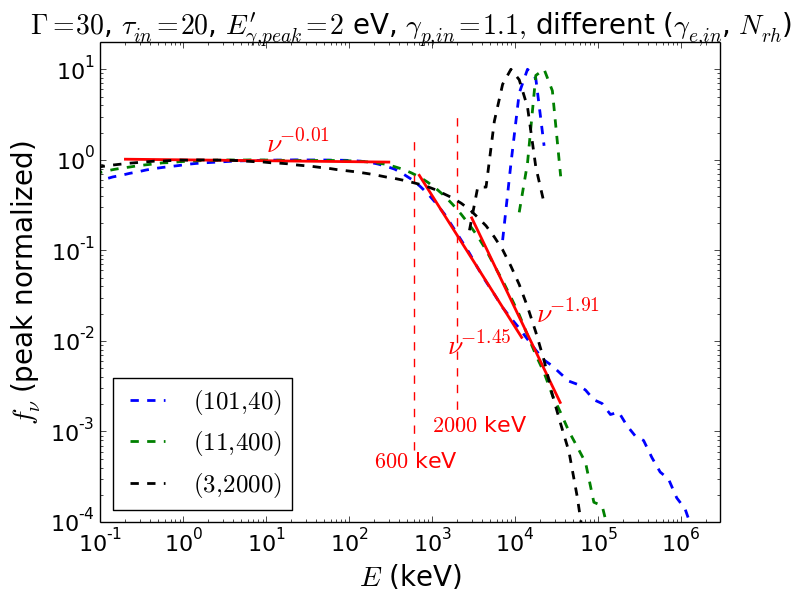}}
  \hspace*{1pt}
  \parbox{2.4in}{\includegraphics[width=2.4in]{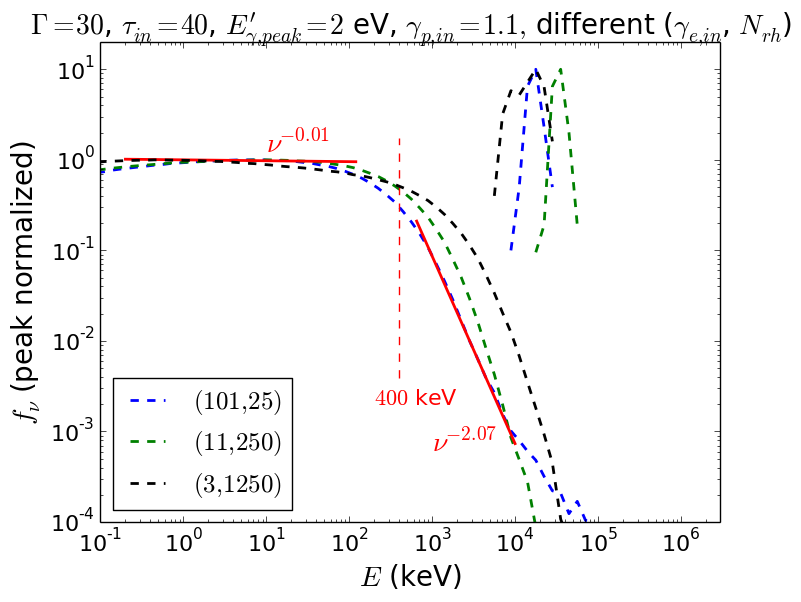}}
  \caption{\textbf{Effect of $\gamma_{e,\rm in}$ at constant $E_{\rm inj}=E_{\rm inj,cr}(\tau_{\rm in})$ and for different $\tau_{\rm in}$}. The input parameters used are $E_{\gamma,\rm peak}^{\prime}=2$ eV, $\gamma_{p,\rm in}=1.1$, $L=10^{52}\ {\rm erg/s}$ and $\Gamma=30$. 
	{\it Left panel:} $\tau_{\rm in}=20$, $E_{\rm inj} = 4000\ m_e c^2$ and $(N_{\rm rh},\gamma_{e,\rm in}) = (40,101)/(400,11)/(2000,3)$.
	{\it Right panel:} $\tau_{\rm in}=40$, $E_{\rm inj} = 2500\ m_e c^2$ and $(N_{\rm rh},\gamma_{e,\rm in}) = (25,101)/(250,11)/(1250,3)$.}%
 \label{fig2}
\end{center}
\end{figure}

In Figure \ref{fig2}, we present the simulation results for fixed injected energy $E_{\rm inj,cr} = N_{\rm rh,cr}(\gamma_{e,in}-1)\ m_e c^2 = 4000/2500\ {\rm m_e c^2}$ at $\tau_{in}=20/40$ and different $\gamma_{e,in}=3,11,101$. The photons/protons are initialized with $E_{\gamma,peak}^{\prime}=2$ eV/$\gamma_{p,in}=1.1$ with jet parameters, $L=10^{52}\ {\rm erg/s}$ and $\Gamma=30$. We find that $\alpha \sim 0$ is practically unaffected by decrease in electron energy $\gamma_{e,in}$ (irrespective of $\tau_{in}$) and is solely determined by the critical injected energy $E_{inj,cr}$. As $\gamma_{e,in}$ increases for a given $E_{inj,cr}$, the photons tend to have lower peak energy $E_{\gamma,peak}$ and there are fewer/more photons with $E_{\gamma} \sim 1-10\ {\rm MeV}$/$\gtrsim 100\ {\rm MeV}$. This is expected as the electrons with $\gamma_{e,in} = 101$ are accelerated much less frequently compared to those with $\gamma_{e,in} \sim 3-11$ and then subsequently cool down very rapidly to non-relativistic $\gamma_{e}$ after being considerably hotter for a shorter duration $\sim 10^{-3}\ t_{dyn}$ when they accelerate many photons to $E_{\gamma} \gtrsim 100\ {\rm MeV}$. The high energy bump in $f_{\nu}$ and deviation from power-law behaviour for large $\gamma_{e,in}$ is seen only at moderate $\tau_{in} \lesssim 20$ and is not appreciable for larger $\tau_{in} \gtrsim 40$ as the high energy photons cool down rapidly from adiabatic losses.

\begin{figure}[h]%
\begin{center}
  \parbox{2.4in}{\includegraphics[width=2.4in]{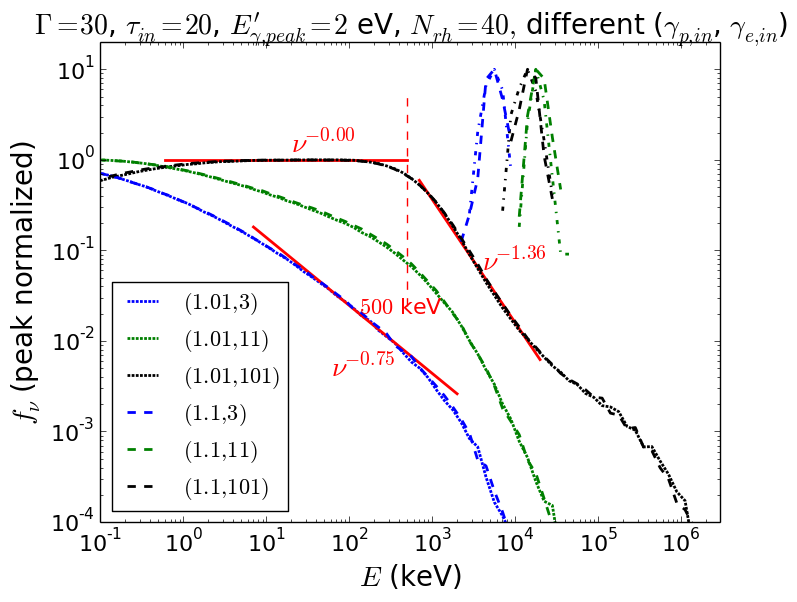}}
  \hspace*{1pt}
  \parbox{2.4in}{\includegraphics[width=2.4in]{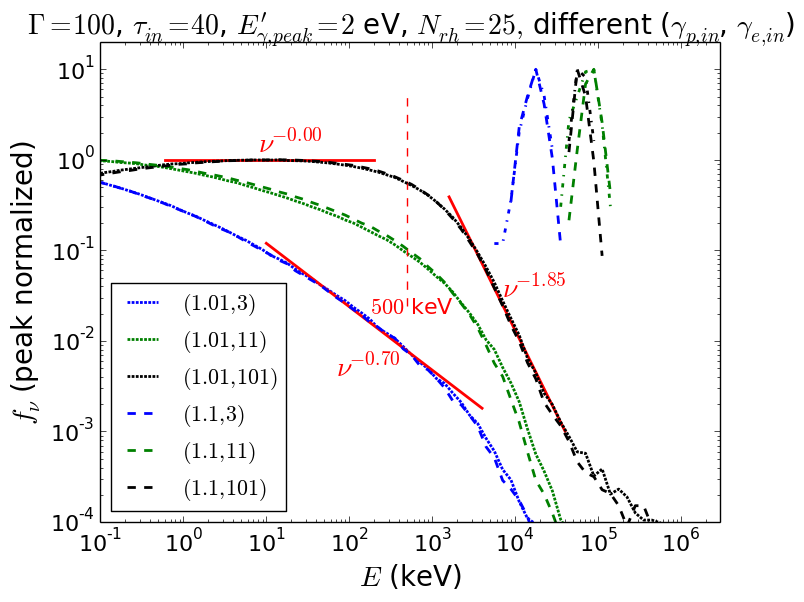}}
  \caption{\textbf{Effect of $\gamma_{p,in}$ and $\gamma_{e,in}$ for $N_{rh}=40\ (\tau_{in}=20, \Gamma=30)$ and $N_{rh}=25\ (\tau_{in}=40, \Gamma=100)$}. The input parameters used are $E_{\gamma,peak}^{\prime}=2$ eV and $L=10^{51}$ erg/s. 
	{\it Left panel:} $\Gamma=30$, $\tau_{in}=20$, $N_{rh}=40$: 
	$\gamma_{p,in}=(1.01,1.1)$ and $\gamma_{e,in}=(3,11,101)$. 	
	{\it Right panel:} $\Gamma=100$, $\tau_{in}=40$, $N_{rh}=25$: 
	$\gamma_{p,in}=(1.01,1.1)$ and $\gamma_{e,in}=(3,11,101)$. 
	}%
 \label{fig3}
\end{center}
\end{figure}

We find an increase in $|\beta|$ with decrease in $\gamma_{e,in}$ for a fixed $E_{inj,cr}(\tau_{in})$ as well as with increase in $\tau_{in}$. Moreover, $\beta \sim \beta_{obs}$ for $\gamma_{e,in} \sim {\rm few\ 10{\rm s}}-100$ and $\tau_{in} \lesssim 20$ while the high-energy spectrum is much steeper, $f_{\nu} \propto \nu^{-2}$ for $\tau_{in} \gtrsim 40$, almost independent of $N_{rh}$. 
While relatively continuous energy injection (small $\gamma_{e,in} \sim {\rm few}$ and large $N_{rh,cr} \sim {\rm few}\ 1000{\rm s}$) results in steeper high energy spectra $|\beta|>|\beta_{obs}|$ along with $E_{\gamma,peak}/E_{\gamma,obs} \gtrsim 10$, episodic energy injection (large $\gamma_{e,in}\sim 100$ and small $N_{rh,cr} \sim {\rm few}\ 10{\rm s}$) gives a high energy power-law spectrum consistent with observations for moderate optical depths $\tau_{in} \lesssim 20$. In order to have both $E_{\gamma,peak} \sim 500\ {\rm keV}$ and $|\beta| \sim 1.2-1.5$, the particles and photons have to be initialized at $\tau_{in} \sim 20-40$ and $E_{inj,cr}(\tau_{in}) \sim 2500-4000\ m_e c^2$ energy needs to be injected into electrons with $\gamma_{e,in} \sim {\rm few}\ 10{\rm s}$.

In Figure \ref{fig3}, we present the simulation results for fixed $N_{\rm rh,cr}(\tau_{\rm in})=40/25$ at $\tau_{\rm in}=20\ (\Gamma=30)/40\ (\Gamma=100)$ with different combinations of $\gamma_{e,\rm in}=3,11,101$ and $\gamma_{p,\rm in}=1.01,1.1$. The seed photons have energy $E_{\gamma,\rm peak}^{\prime}=2\ {\rm eV}$ with jet luminosity $L = 10^{51}\ {\rm erg/s}$. We see that $\gamma_{p,\rm in}$ does not affect the photon output spectra irrespective of the optical depth, which is expected as the Coulomb collisions timescale $t_{e-p}^{\prime} = (\gamma_{e,avg} - 1)m_e c^2/\dot{E}_{e-p}$ is considerably longer than the Comptonization timescale $t_{IC}^{\prime}$. A minimum electron energy $\gamma_{e} \gtrsim 11$ is needed in order to have photons with $E_{\gamma} \gtrsim 10\ {\rm MeV}$ and peak energy $E_{\gamma,peak} \sim 1\ {\rm MeV}$ for both $\tau_{in}$ considered. The output photon spectrum does not show a power-law dependence at both low and high energies when the electron initial energy is small $\gamma_{e,in} \lesssim 11$. While the output photon spectrum shows $\alpha \sim \alpha_{obs}$ and $E_{\gamma,peak} \sim E_{\gamma,obs}$ at both optical depths for electrons with $\gamma_{e,in}=101$ only, the high energy power-law spectral index $|\beta| \gg |\beta_{obs}|$ for $\tau_{in}=40$ and $\sim |\beta_{obs}|$ for $\tau_{in}=20$.

\begin{figure}[h]%
\begin{center}
  \parbox{2.4in}{\includegraphics[width=2.4in]{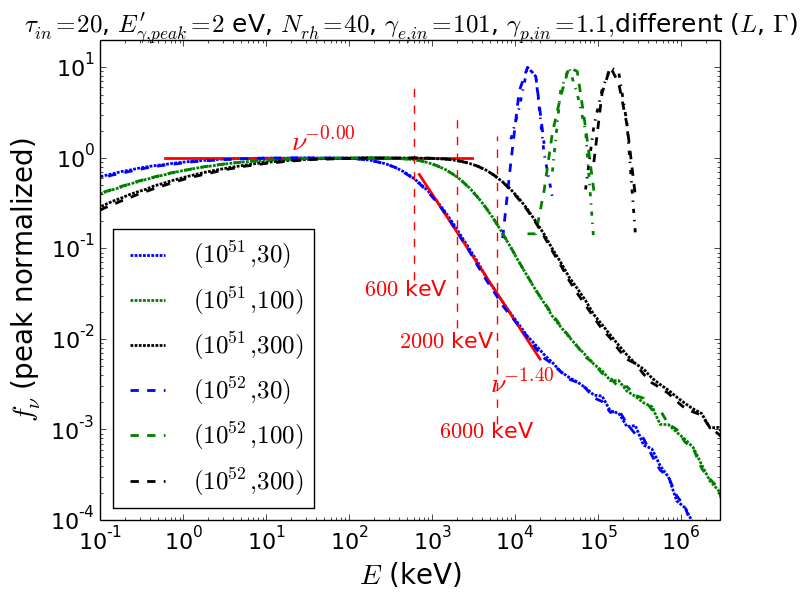}}
  \hspace*{1pt}
  \parbox{2.4in}{\includegraphics[width=2.4in]{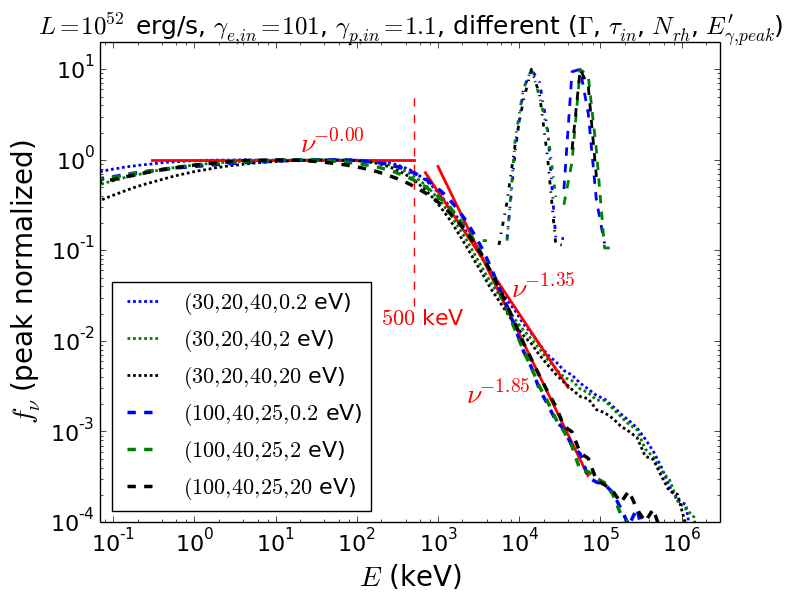}}
  \caption{{\it Left panel:} MCRaT simulations showing the effect of $L=(10^{51},10^{52})\ {\rm erg/s}$ and $\Gamma=30,100,300$ for constant $N_{rh}=40$, $\gamma_{e,in}=101$ and $\tau_{in}=20$. For these simulations, we consider input parameters $E_{\gamma,peak}^{\prime}=2$ eV and $\gamma_{p,in}=1.1$.	 
	{\it Right panel:} MCRaT simulations showing the effect of $E_{\gamma,peak}^{\prime}=0.2,2,20\ {\rm eV}$ for $N_{rh}=40\ (\tau_{in}=20, \Gamma=30)$ and $N_{rh}=25\ (\tau_{in}=40, \Gamma=100)$. For these simulations, we consider input parameters $\gamma_{e,in}=101$, $\gamma_{p,in}=1.1$ and $L=10^{52}\ {\rm erg/s}$.  
	}%
 \label{fig4}
\end{center}
\end{figure}

In the left panel of Figure \ref{fig4}, we show the simulation results for fixed $E_{\rm inj,cr} = 4000\ m_e c^2$ at $\tau_{in}=20$ and for different combinations of $L = 10^{51}, 10^{52}\ {\rm erg/s}$ and $\Gamma = 30, 100, 300$. The photons/electrons/protons are initialized with energies $E_{\gamma,peak}^{\prime}=2\ {\rm eV}/\gamma_{e,in}=101/\gamma_{p,in}=1.1$ at optical depth $\tau_{in}=20$. While the jet luminosity $L$ has no noticeable effect on the output photon spectrum, increase in bulk Lorentz factor $\Gamma$ shifts the photon peak energy to higher values. We find that even though $\Gamma$ does not affect $\alpha$ and $\beta$, it rescales photon peak energy as $E_{\gamma,peak} \propto \Gamma$. The output photon spectrum shows $E_{\gamma,peak} \sim E_{\gamma,obs}$ only for smaller $\Gamma \sim 30$ values. While larger $\Gamma \sim 100$ can also reproduce $E_{\gamma,peak} \sim 500\ {\rm keV}$ and $\alpha \sim 0$ at $\tau_{in} \gtrsim 40$ in agreement with the observations, it cannot explain the observed high energy spectral index (see right panel of Figure \ref{fig3}). 

In the right panel of Figure \ref{fig4}, we show the simulation results for fixed $N_{rh,cr}(\tau_{in})=40/25$ at $\tau_{in}=20\ (\Gamma=30)/40\ (\Gamma=100)$ and different seed photon energies $E_{\gamma,peak}^{\prime}=0.2,2,20\ {\rm eV}$. The electrons/protons are initialized with energies $\gamma_{e,in}=101/\gamma_{p,in}=1.1$ with jet luminosity $L=10^{52}\ {\rm erg/s}$. We find that for both $E_{inj,cr}(\tau_{in})=2500\ m_e c^2$ and $4000\ m_e c^2$, the low/high energy spectral index $\alpha/\beta$ and the output photon peak energy $E_{\gamma,peak}$ are practically unaffected by the choice of $E_{\gamma,peak}^{\prime}$. 
The specific photon flux $f_{\nu}$ falls off considerably at energies less than $\Gamma E_{\gamma,peak}^{\prime}$ as most of the photons gain energy and do not populate the low energy tail after getting scattered by the electrons. For larger $\tau_{in}$, the photons get scattered multiple times thereby increasing the probability of differential number of scatterings before escaping the photosphere and subsequent broadening of the spectrum. As a result, more photons populate the low energy tail and the spectra with different initial energies $\Gamma E_{\gamma,peak}^{\prime}$ become indistinguishable for $\tau_{in} \gtrsim 40$. 

\section{Photon spectra for repeated scatterings}
Assuming that Comptonization is the dominant process influencing the output photon spectrum, we will first evaluate the energy spectrum of synchrotron photons after they experience single scattering with the electrons. Then we extend our formalism to find the photon energy spectrum for the realistic case when they undergo repeated scatterings with the electrons in the jet before exiting the photosphere. The energy distribution of the scattered photons depends mainly on the incident photon spectrum and the electron energy distribution.

\subsection{Photon distribution after one scattering}
For our calculation, we consider electrons and incident photons with isotropic distributions in the jet-comoving frame. In this case, the scattered photons are also distributed isotropically in the jet-comoving frame. For simplicity, we only consider Thomson scattering in the rest frame of the electron and assume that all scattering events are elastic in nature. For incident photons with energy $\epsilon$ scattering off electrons with energy $\gamma m_e c^2$, the total scattered power per energy per volume is \cite{RL79}
\begin{equation}
\frac{dE}{dVdtd\epsilon_{1}} = \frac{3}{4}c\sigma_{T}\int_{\epsilon_{1}/4\gamma^2}^{\infty} d\epsilon \frac{\epsilon_{1}}{\epsilon^{2}}f(\epsilon)\int_{1}^{\infty} \frac{d\gamma}{\gamma^{2}}n_{e}(\gamma)g_{\rm iso}\left(\frac{\epsilon_{1}}{4\gamma^2 \epsilon}\right),
\label{out_sc}
\end{equation}
where, $\epsilon_{1}$ is the scattered photon energy, $f(\epsilon)$ is the photon distribution function, $n_{e}(\gamma)$ is the electron distribution function and $g_{\rm iso}(x) = \frac{2}{3}(1-x)$ for isotropic photon distribution in the jet-comoving frame.
We consider the simple case in which the incident photons have a synchrotron/piecewise power-law energy distribution,
\begin{equation}
f_{in}(\epsilon) = f_{0}\left\{
\begin{array}{ll}
(\epsilon/\epsilon_{0})^{a}, & \epsilon<\epsilon_{0}\\
(\epsilon/\epsilon_{0})^{-b}, & \epsilon>\epsilon_{0}\\
\end{array}
\right. 
\label{FCphseed}
\end{equation}
and the electrons are mono-energetic with $n_{e}(\gamma) = n_{0}\delta(\gamma - \gamma_{0})$. Substituting $x=4\gamma_{0}^{2}\epsilon/\epsilon_{1}$  for relativistic electrons yields
\begin{equation}
\frac{dE}{dVdtd\epsilon_{1}} = 2c\sigma_{T}n_{0}\int_{1}^{\infty}\frac{dx}{x^2}\left(1 - \frac{1}{x}\right)f\left(\frac{\epsilon_{1}x}{4\gamma_{0}^{2}}\right).
\end{equation}

For photons below peak energy $\epsilon_{1} < 4\gamma_{0}^{2}\epsilon_{0}$, we can further define $\epsilon_{1}/4\gamma_{0}^{2} = \eta \epsilon_{0}$ with $\eta < 1$. For low energy photons with $\eta \ll 1$, if the incident photons have a hard spectrum with $0 < a < 1$, the scattered photon distribution $f_{sc}(\epsilon_{1}) \propto dE/(dVdtd\epsilon_{1}) \propto \eta^{a} \propto \epsilon_{1}^{a}/(\gamma_{0}^{2a}\epsilon_{0}^{a})$ is the same as that of the incident photons. However, for a softer low energy incident photon spectrum with $a \ge 1$ and $\eta \ll 1$, we obtain $f_{sc}(\epsilon_{1}) \propto dE/(dVdtd\epsilon_{1}) \propto \epsilon_{1}$. Therefore, after single scattering of synchrotron photons with broken power-law energy distribution, the low energy spectrum is unaffected for hard spectra with $a<1$ whereas $f_{sc}(\epsilon) \propto \epsilon$ for softer spectra. 
For photons above peak energy $\epsilon_{1} > 4\gamma_{0}^{2}\epsilon_{0}$ and $\eta \gtrsim 1$. This gives $dE/(dVdtd\epsilon_{1}) \propto \epsilon_{1}^{-b}/(\gamma_{0}^{-2b}\epsilon_{0}^{-b})$, same as the incident photon spectrum. 
For fast cooled synchrotron photon spectrum, we have $a=2$ and $b=-1$, and the scattered photon distribution after single scattering is
\begin{equation}
f_{1}(\epsilon) = f_{sc}(\epsilon) \propto f_{0}\left\{
\begin{array}{ll}
(\epsilon/\epsilon_{0})^{1}, & \epsilon<\epsilon_{0}\\
(\epsilon/\epsilon_{0})^{-1}, & \epsilon>\epsilon_{0}\\
\end{array}
\right. 
\end{equation}
In reality, however, each photon experiences $\sim 2\tau_{in}$ scatterings on an average before escaping the photosphere.

\subsection{Photon distribution after repeated scatterings}
With $f_{1}(\epsilon)$ as the incident photon distribution, we can now extend the same formalism to calculate the photon spectrum after subsequent scattering events assuming that the electron and photon distributions remain isotropic in the jet-comoving frame. The low and the high energy spectrum after each photon in the jet has undergone exactly two scatterings is
\begin{eqnarray}
f_{2,l}(\epsilon_1) = 2c\sigma_{T}n_{0}f_{0}\left(-\eta{\rm ln}\eta + \frac{2}{3}\eta^{2} - \frac{1}{2}\eta\right)
\propto \frac{\epsilon_{1}}{4\gamma_{0}^{2}\epsilon_{0}}{\rm ln}\left(\frac{\epsilon_{1}}{4\gamma_{0}^{2}\epsilon_{0}}\right),\nonumber \\
f_{2,u}(\epsilon_1) = 2c\sigma_{T}n_{0}f_{0}\int_{1}^{\infty}dx\frac{1}{x^2}\left(1 - \frac{1}{x}\right)\eta^{-1}x^{-1} \propto \gamma_{0}^{2}\epsilon_{0}/\epsilon_{1}. \nonumber
\end{eqnarray}
The scattered photon spectrum is then
\begin{equation}
f_{2}(\epsilon) \propto f_{0}\left\{
\begin{array}{ll}
(\epsilon/\epsilon_{0}){\rm ln}(\epsilon/\epsilon_{0}), & \epsilon<\epsilon_{0}\\
(\epsilon/\epsilon_{0})^{-1}, & \epsilon>\epsilon_{0}\\
\end{array}
\right. 
\end{equation}
After each photon has undergone exactly three scatterings, the low and high energy spectra are given as
\begin{eqnarray}
f_{3,l}(\epsilon_1) \propto \frac{\epsilon_{1}}{4\gamma_{0}^{2}\epsilon_{0}}\left[{\rm ln}\left(\frac{\epsilon_{1}}{4\gamma_{0}^{2}\epsilon_{0}}\right)\right]^{2}, \nonumber \\
f_{3,u}(\epsilon_1) = 2c\sigma_{T}n_{0}f_{0}\int_{1}^{\infty}dx\frac{1}{x^2}\left(1 - \frac{1}{x}\right)\eta^{-1}x^{-1}
\propto \gamma_{0}^{2}\epsilon_{0}/\epsilon_{1}, \nonumber
\end{eqnarray}
and the scattered photon spectrum is
\begin{equation}
f_{3}(\epsilon) \propto f_{0}\left\{
\begin{array}{ll}
(\epsilon/\epsilon_{0})\left[{\rm ln}(\epsilon/\epsilon_{0})\right]^{2}, & \epsilon<\epsilon_{0}\\
(\epsilon/\epsilon_{0})^{-1}, & \epsilon>\epsilon_{0}\\
\end{array}
\right. 
\end{equation}
We can generalize the above results further for $N \sim 2\tau_{in}$ scatterings per photon 
\begin{equation}
f_{N}(\epsilon) \propto f_{0}\left\{
\begin{array}{ll}
(\epsilon/4\gamma_{0}^{2}\epsilon_{0})\left[{\rm ln}(\epsilon/4\gamma_{0}^{2}\epsilon_{0})\right]^{N-1}, & \epsilon<\epsilon_{0}\\
(\epsilon/4\gamma_{0}^{2}\epsilon_{0})^{-1}, & \epsilon>\epsilon_{0}\\
\end{array}
\right. 
\label{sc_N}
\end{equation}

\begin{figure}[h]%
\begin{center}
  \parbox{2.4in}{\includegraphics[width=2.4in]{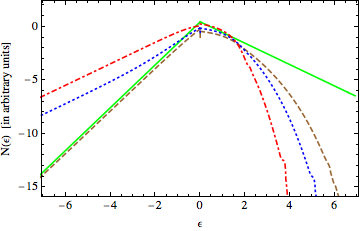}}
  \hspace*{1pt}
  \parbox{2.4in}{\includegraphics[width=2.4in]{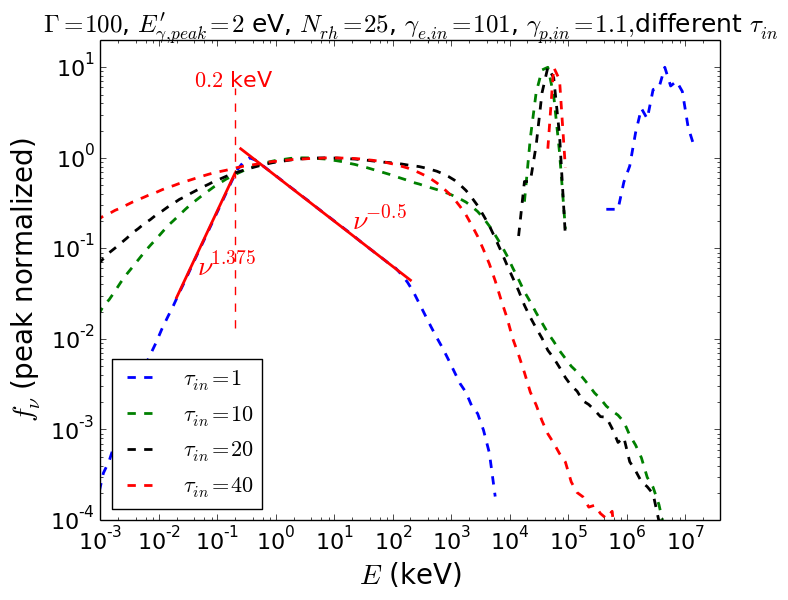}}
  \caption{\textbf{Effect of geometrical broadening on the photon spectrum for increasing optical depth.}
  {\it Left panel:} IC spectrum for synchrotrom photons with energy $\epsilon_0=1$ and Maxwellian electrons with peak energy $\gamma_{e,0} = 1.1$. The solid green line, brown dashed line, blue dotted line and red dot-dashed lines are the scattered photon spectra after $N = 0, 1, 2$ and $5$ scatterings, respectively.  
  {\it Right panel:} MCRaT simulations showing the effect of geometrical broadening on the photon spectrum at $\tau_{in}=1,10,20,40$ for $(N_{rh},\gamma_{e,in})=(25,101)$ and $\gamma_{p,in}=1.1$. For these simulations, we consider input parameters $L=10^{52}\ {\rm erg/s}$, $E_{\gamma,peak}^{\prime} = 2\ {\rm eV}$ and $\Gamma=100$.  
	}%
 \label{fig5}
\end{center}
\end{figure}

In Figure \ref{fig5}, we show how the photon spectrum is affected by Comptonization with electrons as $\tau_{in}$ and number of scatterings increase. In the left panel, the IC scattered photon spectrum for fast cooled synchrotron seed photons (Equation \ref{FCphseed}, with $a=2$ and $b=-1$) with energy $\epsilon_0 = m_e c^2$ and Maxwellian electrons with peak energy $\gamma_{e,0}=1.1$ is shown for scattering orders $N=0,1,2,5$. As predicted by equation (\ref{sc_N}), the photon spectrum becomes gradually softer below peak energy as the scattering order increases. The photons scattering off Maxwellian electrons with fixed energy get thermalized at equilibrium to attain a high-energy exponential tail for large optical depths/scatterings. Furthermore, there is gradual flattening of the low-energy spectrum with increase in scattering order $N$. 

In the right panel of Figure \ref{fig5}, we present the MCRaT simulation results for $E_{\rm inj,cr} = 2500\ m_e c^2$ and different optical depths $\tau_{in}=1,10,20,40$. The number of repeated dissipation events in the jet are $N_{rh}=25$ with initial photon/electron/proton energy $E_{\gamma,peak}^{\prime}=2\ {\rm eV}$/$\gamma_{e,in} = 101$/$\gamma_{p,in} = 1.1$ and jet parameters $L=10^{52}\ {\rm erg/s}$ and $\Gamma=100$. With increase in scattering order ($\propto \tau_{in}$), the high energy photon spectrum becomes steeper with a simultaneous decrease in $E_{\gamma,peak}$. These photons then populate the low energy spectrum and extend the non-thermal tail to energies much lower than $E_{\gamma,peak} \sim 0.2\ {\rm keV}$. The photon spectra from simulations are also considerably broader compared to the analytical results for similar values of $N$.

\section{Summary \& Conclusions}
We have utilized our MCRaT photospheric code to explain the distinct non-thermal behaviour of GRB prompt emission spectrum, $f_{\nu} \propto \nu^{0}/f_{\nu} \propto \nu^{-1.2}$ at low/high photon energies along with observed peak energy at $E_{\gamma,peak} \sim 300\ {\rm MeV}$. For our simulations, we considered Comptonization of fast cooled synchrotron photons with Maxwellian electrons and for photon to particle number ratio $N_{\gamma}/N_{e} \sim 10^5$. The jet electrons are accelerated by two different mechanisms: 1. continuous energy transfer via Coulomb collisions with mono-energetic protons, 2. repeated episodic energy dissipation events that are equally spaced over time and accelerate particles back to their initial energies. 

\begin{figure}[h]%
\begin{center}
 \parbox{3.2in}{\includegraphics[width=3.2in]{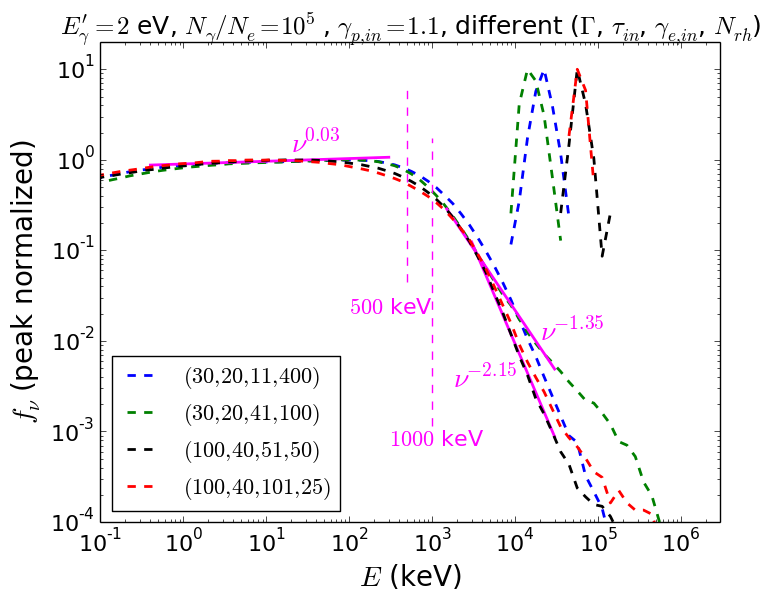}}
\caption{\textbf{MCRaT simulation results with the best set of parameters for a jet with $N_{\gamma}/N_{e}=10^5$.}
The relativistic jet with $L=10^{52}\ {\rm erg/s}$ has photons with $E_{\gamma,\rm peak}^{\prime}=2\ {\rm eV}$ and protons with $\gamma_{p,\rm in}=1.1$. The energy injection necessary in order to produce an output photon spectrum with the observed Band-like spectral properties depends on $\tau_{\rm in}$ and $\Gamma$. Here we consider $E_{\rm inj}=4000/2500\ m_e c^2$ for $\Gamma=30/100$ and $\tau_{\rm in}=20/40$, for two distinct electron energies $\gamma_{e,\rm in}=(11,41)/(51,101)$.   
}%
 \label{fig7}
\end{center}
\end{figure}

The parameters that significantly affect the spectral properties for a given $N_{\gamma}/N_{e}$ are $E_{\rm inj}(\gamma_{e,\rm in},N_{\rm rh})$, $\Gamma$ and $\tau_{\rm in}$. Figure \ref{fig7} shows the most probable parameter set that gives output photon spectrum with $(\alpha,\beta,E_{\gamma,\rm peak})$ very similar to the observed GRB prompt spectrum. The photons/protons in these simulations are initialized with energies $E_{\gamma,\rm peak}^{\prime}=2\ {\rm eV}$/$\gamma_{p,\rm in}=1.1$ for jet parameters $L = 10^{52}\ {\rm erg/s}$, $N_{\gamma}/N_{e}=10^5$ and $\Gamma \sim 30-100$. The particles are injected with energy $E_{\rm inj,cr} \sim 2500-4000\ m_e c^2$ per electron for a range of $\tau_{\rm in} \sim 20-40$. For smaller $\tau_{in} \sim 20$ and $\Gamma \sim 30$, $(\alpha,\beta,E_{\gamma,peak}) \sim (0,-1.4,1\ {\rm MeV})$ is obtained with $E_{inj,cr} \sim 4000\ m_e c^2$ and $\gamma_{e,in} \gtrsim 40$. Although $\alpha \sim 0$ and $E_{\gamma,peak} \sim 500\ {\rm keV}$ for $E_{inj,cr} \sim 2500\ m_e c^2$ at larger $\tau_{in} \sim 40$ and $\Gamma \sim 100$, the high energy spectrum is significantly steeper than the observed prompt spectrum with $\beta \sim -2.1$, especially for $\gamma_{e,in} \lesssim 50$. 

The main results of this work can be summarised as: 
\begin{enumerate}
\item The electrons cool down rapidly to non-relativistic energies in the absence of jet dissipation events. This entails energy injection into the jet particles via either (continuous) Coulomb collisions or (episodic) sub-photospheric dissipation events. However, for Coulomb heating, the protons lose a considerable fraction of their energy within $\sim t_{dyn}$, for $\tau_{in} \gtrsim 10$, to attain non-relativistic energies. Therefore, continuous energy injection by protons is not sufficient to maintain electrons at $\gamma_{e} \sim \gamma_{e,crit}$. 

\item Energy injection can be achieved with episodic sub-photospheric dissipation events which can keep the electrons at energies $\gamma_{e} \gtrsim \gamma_{e,crit}$ provided that they are sufficiently energetic and frequent. For large $E_{\rm inj}$, the photon peak energy $E_{\gamma,\rm peak} \gg E_{\gamma,\rm obs} \sim 300\ {\rm keV}$, while for large $\tau_{\rm in}$, $E_{\gamma,\rm peak} \ll E_{\gamma,\rm obs}$ due to significant adiabatic loss. From MCRaT simulations, we quantify the $E_{inj} - \tau_{\rm in}$ correlation: injected energy $E_{inj,cr} = 4000/2500\ m_e c^2$ per electron for initial optical depth $\tau_{in} = 20/40$. 

\item In the output photon spectrum, $\alpha$ critically depends on $E_{\rm inj}$ whereas $\beta$ and $E_{\gamma,\rm peak}$ are almost entirely determined by $\tau_{\rm in}$. With an increase in $\tau_{\rm in}$, $E_{\gamma,\rm peak}$ decreases and the high-energy photon spectrum becomes steeper. Additionally, $|\beta|$ also increases with decrease in initial electron energy $\gamma_{e,in}$ for fixed $E_{\rm inj,cr} = N_{rh,cr}(\gamma_{e,in}-1)\ m_e c^2$. We find that $E_{\gamma,\rm peak} \sim E_{\gamma,\rm obs}$ only for smaller $\Gamma \sim 30$ - while larger $\Gamma \sim 100$ gives $E_{\gamma,peak} \sim 500\ {\rm keV}$ at $\tau_{in} \sim 40$, the high energy photon spectrum is considerably steeper than observed.

\item For isotropic electrons scattering isotropic photons, the scattered photon energy distribution can be analytically evaluated for lower order scatterings, given the electron and photon energy distributions. For Comptonization of synchrotron photons with Maxwellian electrons, $\alpha \sim 0$ behaviour is retained at low energies whereas $f_{\nu} \propto e^{-\nu}$ at high energies. Qualitatively, the low-energy non-thermal dependence is obtained from multiple scatterings and subsequent geometrical broadening of the spectrum whereas the high-energy power-law dependence is attributed to repeated episodic and continuous energy injection events.
\end{enumerate}

\end{document}